\documentclass[final]{aipproc}

\layoutstyle{8x11single}

\usepackage{amsmath}
\begin{document}

\title{Structure and prospects of the simplest $SO(10)$ GUTs}


\author{Michal Malinsk\'{y}\footnote{Presenting author; email:malinsky@ific.uv.es}\hskip 2mm}{
  address={AHEP Group, Instituto de F\'{\i}sica Corpuscular -- C.S.I.C./Universitat de Val\`encia, Edificio de Institutos de Paterna, Apartado 22085, E 46071 Val\`encia, Spain}
}

\author{Stefano Bertolini}{
  address={INFN, Sezione di Trieste, SISSA,
via Bonomea 265, 34136 Trieste, Italy}
}

\author{Luca Di Luzio}{
  address={Institut f\"{u}r Theoretische Teilchenphysik,
Karlsruhe Institute of Technology (KIT), D-76128 Karlsruhe, Germany}
}

\begin{abstract}
We recapitulate the latest results on the class of the simplest $SO(10)$ grand unified models in which the GUT-scale symmetry breaking is triggered by an adjoint Higgs representation. We argue that the minimal survival approximation traditionally used in the GUT- and seesaw-scale estimates tends to be blind to very interesting parts of the parameter space in which some of the intermediate-scale states necessary for non-supersymmetric unification of the SM gauge couplings can be as light as to leave their imprints in the TeV domain.  The stringent minimal-survival-based estimates of the $B-L$ scale are shown to be  relaxed by as much as four orders of magnitude, thus admitting for a consistent implementation of the standard seesaw mechanism even without excessive fine-tuning implied by the previous studies. The prospects of the minimal renormalizable $SO(10)$ GUT as a potential candidate for a well-calculable theory of proton decay are discussed in brief.
\end{abstract}

\maketitle

\section{Introduction}
\begin{figure}[th]
\includegraphics[width=7cm]{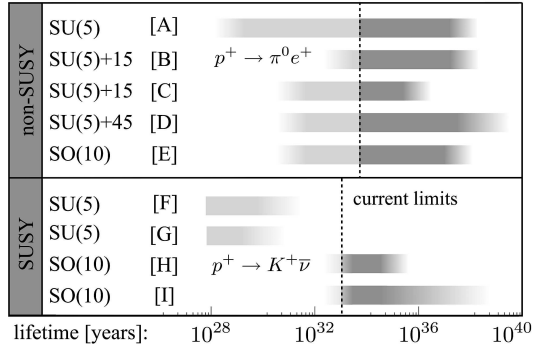}
\caption{\label{estimates} A simple illustration of the typical size of uncertainties in proton lifetime estimates obtained in some of the most popular SU(5) and SO(10) GUTs. Rows [A]-[D] depict the results obtained in the Georgi-Glashow model and some of its simplest extensions \cite{Georgi:1974yf,Goldman:1980ah}, \cite{Dorsner:2005fq}, \cite{Dorsner:2006hw} and \cite{Dorsner:2006dj}; in row [E] we quote the range given in \cite{Lee:1994vp}; in the SUSY case the [F] and [H] correspond  to the estimates given in \cite{Pati:2005yp}; finally, [G] and [I] refer to studies \cite{Murayama:2001ur} and \cite{Dutta:2004zh}, respectively. For more information see, e.g., \cite{Nath:2006ut} and references therein.}
\end{figure}

With the next generation of large-volume proton-decay searches and neutrino experiments currently in the R\&D phase (in particular, LBNE~\cite{Akiri:2011dv}, LENA~\cite{Autiero:2007zj} and Hyper-K \cite{Abe:2011ts}) there are good prospects to push the current lower bounds on the proton lifetime to the unprecedented level of $10^{35}$ years. 
On the theory side, the new information may be, at least in principle, used for further testing of the grand unification paradigm; however, this would require a very good grip on the proton lifetime predictions supplied by specific GUTs.  Unfortunately, the quality of the existing estimates is rather limited even in very simple scenarios, see FIGURE~\ref{estimates}, and it is namely due to the low accuracy of the  leading-order methods used in most of the relevant calculations. On the other hand, consistent next-to-leading-order (NLO) proton lifetime estimates are parametrically more difficult:  First, at the NLO level, the GUT scale $M_{G}$ must be determined at two-loops; this, however, requires a detailed understanding of the one-loop theory spectrum. Second, the flavour structure of the relevant baryon-number-violating (BNV) currents must be constrained by the existing data to a maximum attainable degree. Third, one has to account for several classes of almost irreducible uncertainties related to the Planck-scale physics (such as, e.g., gravity smearing of the gauge unification pattern~\cite{Calmet:2008df, Chakrabortty:2008zk}) which are often comparable to the NLO effects. 

Therefore, the only foreseeable way to overcome this conundrum is to focus on the simplest possible GUTs. 
In contrast to the minimal $SU(5)$ Georgi-Glashow model \cite{Georgi:1974sy} which was shown to be incompatible with the electroweak data already back in mid 1980's, the history of the minimal $SO(10)$ GUTs is rather non-linear and even after almost 40 years it is still lively and evolving. Interestingly, this can be partly attributed also to the fact that, in the $SO(10)$ context, the very meaning of minimality is not entirely agreed upon. This owes namely to the relatively large number of potentially viable symmetry breaking chains in $SO(10)$ characterized by different effective scenarios emerging at intermediate scales. Let us recall that this is not the case in SU(5) simply because there the  need to preserve rank reduces the set of Higgs representations available for the GUT symmetry breaking to just few.  

In a certain sense, this is not the case in supersymmetric theories either because the rigidity of the MSSM gauge unification pattern calls for a single-step breaking where most of the details of the GUT-scale dynamics remain obscured. Thus, besides very special features like natural R-parity conservation etc., the main distinctive characteristics of many models is namely their flavour structure.  Hence, with the spectacular failure~\cite{Aulakh:2005mw,Bertolini:2006pe} of the simplest potentially realistic renormalizable SUSY $SO(10)$~\cite{Clark:1982ai,Aulakh:1982sw} (advocated by many to be even the very minimal SUSY GUT~\cite{Aulakh:2003kg}), and, in particular, with no signs of SUSY at the LHC so far, the community's attention naturally drifts back to  non-supersymmetric GUTs. 

In this review, we shall comment in brief on the status of the simplest non-SUSY $SO(10)$ scenarios and on the latest developments including, in particular, the new upper limits on the seesaw scale obtained recently in the work~\cite{Bertolini:2012im} and possible future prospects of accurate proton lifetime calculations in this scenario.

\section{The minimal $SO(10)$ grand unfication}
The simplest multiplet that can consistently support spontaneous breaking of the $SO(10)$ gauge symmetry in the SM direction is the 45-dimensional adjoint representation. Together with either the 16-dimensional spinor or the 126-dimensional self-dual part of the maximally antisymmetric tensor the models based on the combinations $45\oplus 16$ or $45\oplus 126$ are often regarded to as the minimal renormalizable realizations of the Higgs mechanism in the $SO(10)$ GUTs. In this respect, it is important to recall that this is not the case in SUSY where the $F$-flatness conditions align the VEV of $45_{H}$ along that of $16_{H}$ which, although providing the desired rank reduction, leaves a full SU(5) as an unbroken subgroup. Remarkably enough, in the non-supersymmetric Higgs model based on  $45\oplus 16$ or $45\oplus 126$ the SU(5) trap can not be entirely avoided either.  
\subsection{The tree-level curse of the minimal SO(10) GUTs}
The point is that there are two states in the scalar spectrum of either of the two variants of the minimal model that can be simultaneously non-tachyonic only in a narrow region of the parameter space which, unfortunately, happens to support only SU(5)-like symmetry-breaking chains. Indeed, the masses of the color-octet and the $SU(2)_{L}$-triplet components of $45_{H}$ are at the tree level given by \cite{Yasue:1980fy,Anastaze:1983zk,Babu:1984mz}
\begin{eqnarray}
\label{PGBmasses}
M^2(1,3,0)_{45} & = &
2 a_2 (\omega_{BL} - \omega_R) (\omega_{BL} + 2 \omega_R) \, , \\
M^2(8,1,0)_{45} & = &
2 a_2 (\omega_R - \omega_{BL}) (\omega_R + 2 \omega_{BL}) \,,\nonumber
\end{eqnarray}
where $a_{2}$ is a coupling in the relevant scalar potential (see, e.g.,~~\cite{Bertolini:2012im}) and $\omega_{BL}$ and $\omega_{R}$ are the two independent SM-compatible VEVs in $45_{H}$
\begin{equation}
\langle 45_{H} \rangle ={\rm diag}(\omega_{BL},\omega_{BL},\omega_{BL},\omega_{R},\omega_{R})\otimes \tau_{2}
\end{equation} 
(with $\tau_{2}$ denoting the second Pauli matrix) which, if hierarchical enough, break the $SO(10)$ gauge symmetry along two different symmetry breaking chains
\begin{eqnarray}
\label{chainXII}
SO(10)&\stackrel{\omega_{R}}{\longrightarrow}
&SU(4)_C \otimes SU(2)_L \otimes U(1)_R \;  \stackrel{}{\longrightarrow}\ldots \longrightarrow \;  \mbox{SM}\,,\\
\label{chainVIII}
SO(10)&\stackrel{\omega_{BL}}{\longrightarrow}
& SU(3)_c \otimes SU(2)_L \otimes SU(2)_R \otimes U(1)_{B-L} \; \stackrel{}{\longrightarrow}\ldots \longrightarrow \;  \mbox{SM}\,.
\end{eqnarray}
Given that, it is clear that the right hand sides (RHS) of both equations in (\ref{PGBmasses}) are positive if and only if 
$-2\leq\omega_{BL}/\omega_{R}\leq-1/2$,
i.e., when there is essentially no hierarchy between $\omega_{R}$ and $\omega_{BL}$, otherwise the corresponding vacuum is unstable. Obviously, in such a case, $\langle 45_{H} \rangle$ is almost homogeneous and the descent is $SU(5)$-like; this, however, conflicts with the gauge unification constraints as in the Georgi-Glashow model.
\subsection{The quantum salvation}
\begin{figure}[t]
\includegraphics[width=3.5cm]{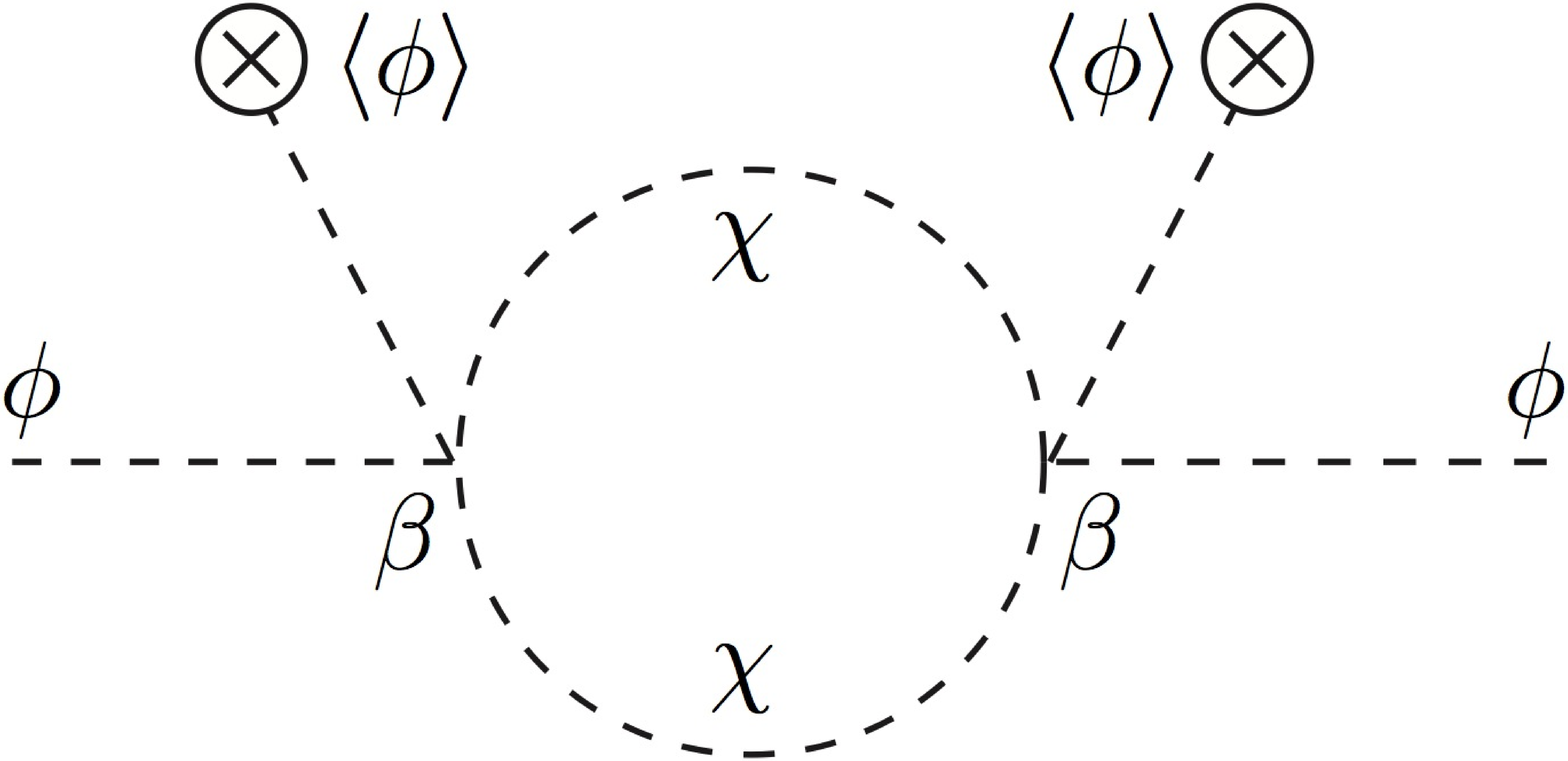}\hskip 10mm
\includegraphics[width=3.5cm]{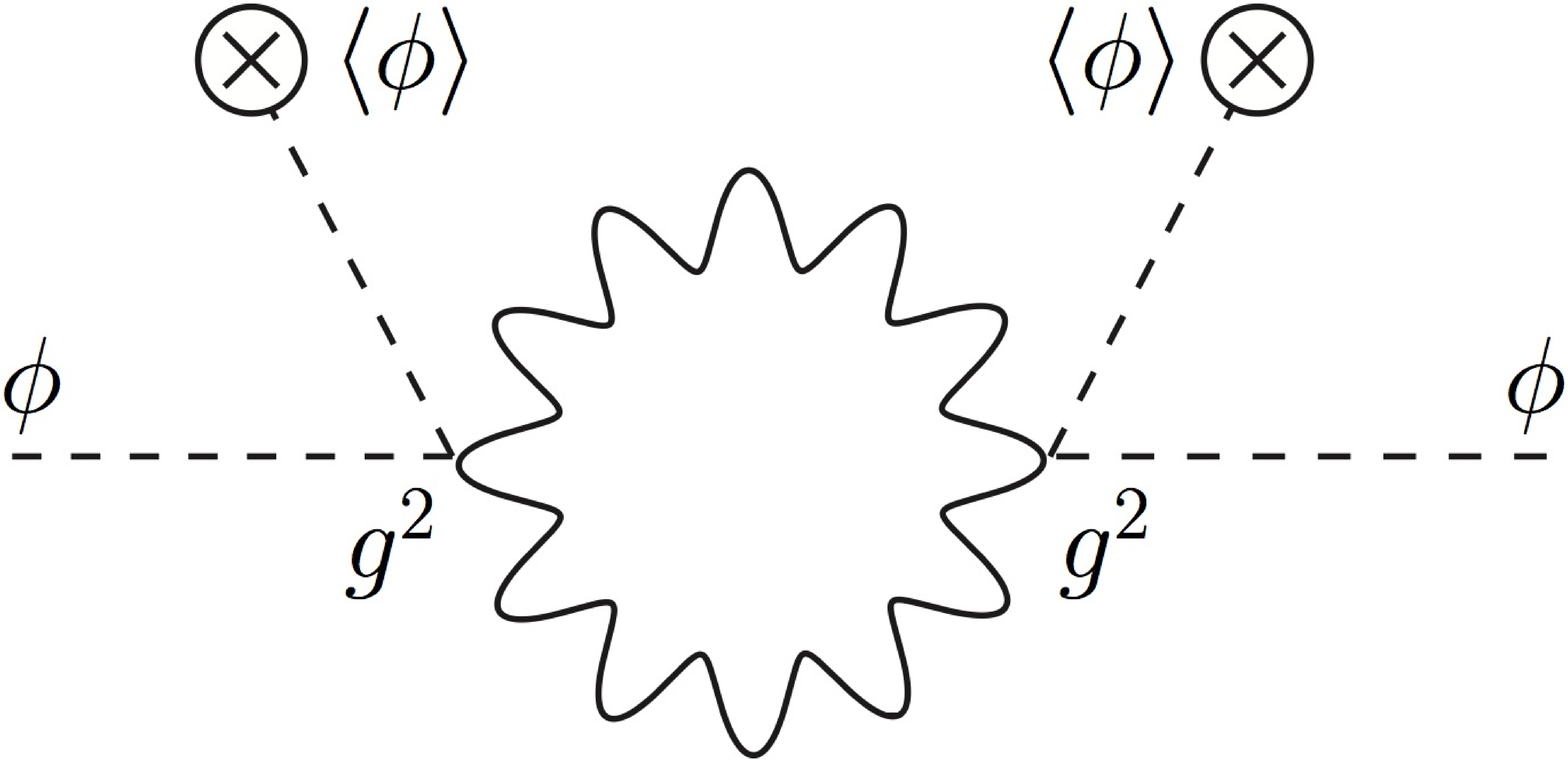}\hskip 10mm
\includegraphics[width=3.5cm]{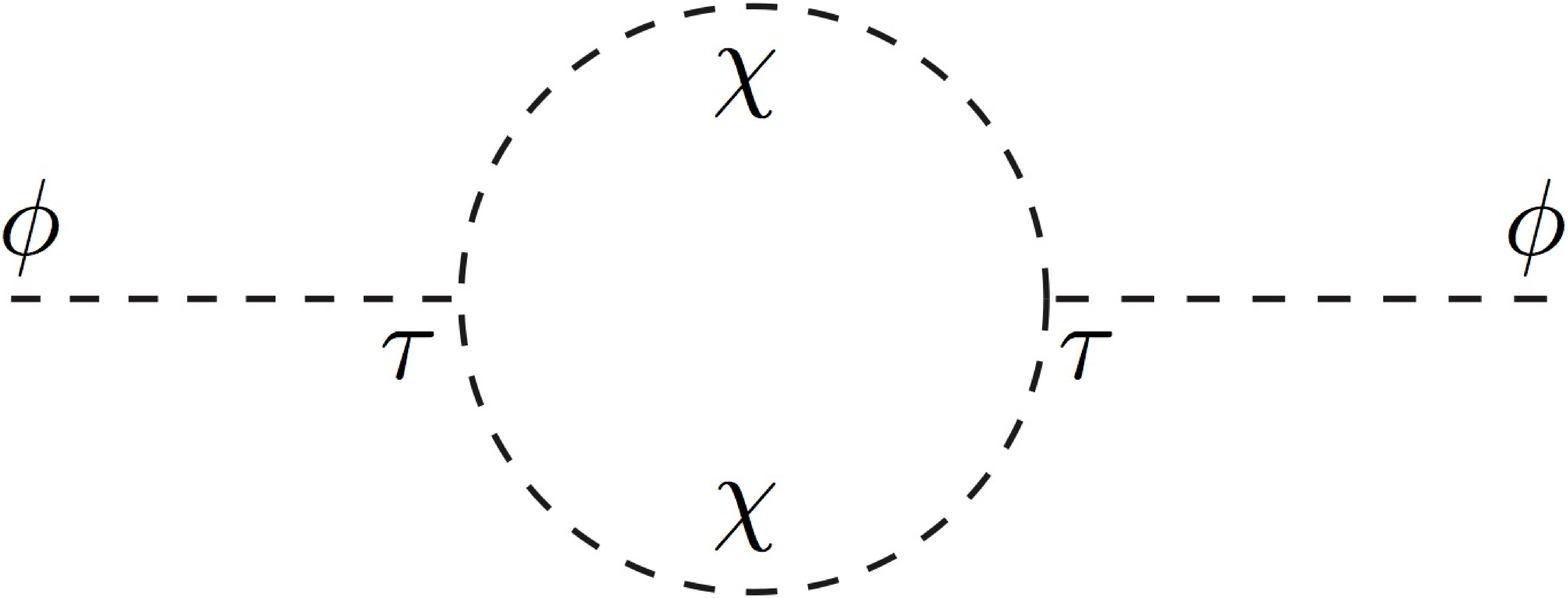}
\caption{\label{loops}Sample topologies of loop diagrams providing the quantum-level stabilization of the potentially realistic vacua in the minimal $SO(10)$ GUT. The generic symbol $\phi$ stands for the components of the 45-dimensional adjoint Higgs representation while $\chi$ denotes components of the complex scalar (either $16_{H}$ or $126_{H}$) responsible for the $B-L$ symmetry breakdown.}
\end{figure}

Until recently, the argument above was taken as a no-go for any potential viability of the minimal $SO(10)$ model including the adjoint scalar as a Higgs field responsible for the initial $SO(10)$ symmetry breakdown. However, as shown in \cite{Bertolini:2009es} this was premature because the hierarchy between $\omega_{BL}$ and $\omega_{R}$ may be stabilized by quantum effects because loop corrections such as those diplayed in FIGURE~\ref{loops} provide non-negligible positive contributions to the RHS of eq.~(\ref{PGBmasses}).
A thorough effective potential analysis~\cite{Bertolini:2009es} in the simplest $45\oplus 16$ variant yields (in the notation of \cite{Bertolini:2009es})
\begin{eqnarray}
\label{310onthevac}
\Delta M^2(1,3,0)_{45}&=& \frac{1}{4\pi^2} \left[ \tau^2
+\beta^2(2\omega_R^2-\omega_R\omega_{BL}+2\omega_{BL}^2)
+g^4 \left(16 \omega_R^2+\omega_{BL} \omega_R+19 \omega_{BL}^2\right)\right]+{\rm logs}\,, \;\;\; 
\nonumber\\
\label{810onthevac}
\Delta M^2(8,1,0)_{45}&=& \frac{1}{4\pi^2} \left[ \tau^2
+\beta^2(\omega_R^2-\omega_R\omega_{BL}+3\omega_{BL}^2)
+g^4 \left(13 \omega_R^2+\omega_{BL} \omega_R+22 \omega_{BL}^2\right)\right]+{\rm logs}\,,\nonumber 
\end{eqnarray}
where the ``logs'' denote the typically sub-leading logarithmic corrections. Hence, for small-enough $a_{2}$ in~(\ref{PGBmasses}) the two problematic states may have non-tachyonic masses even for a large hierarchy between $\omega_{R}$ and $\omega_{BL}$, thus avoiding the tree-level SU(5) trap.
Let us also note that, up to the obvious differences in the ${\cal O}(1)$ factors,  the same dynamical mechanism works in the $45\oplus 126$ Higgs model.
\section{Seesaw scale in the minimal renormalizable $SO(10)$ GUT}
However, the vacuum stability was not the only issue that plagued the $SO(10)$ GUTs for years. The enormous progress in neutrino physics in the last two decades pinned the light neutrino masses into the sub-eV domain with the upper bound (namely, from cosmology and double-beta-decay searches) in the  1 eV ballpark. In the seesaw picture, this typically translates into a lower bound on the scale of the underlying dynamics somewhere in the $10^{12-13}$ GeV domain. This, however, was long ago claimed to be incompatible with the basic features of the symmetry-breaking pattern in the minimal SO(10) models.
\vskip 1mm
\subsection{Seesaw scale in the minimal survival approximation}
\begin{figure}[h]
 \includegraphics[width=6cm, height=5cm]{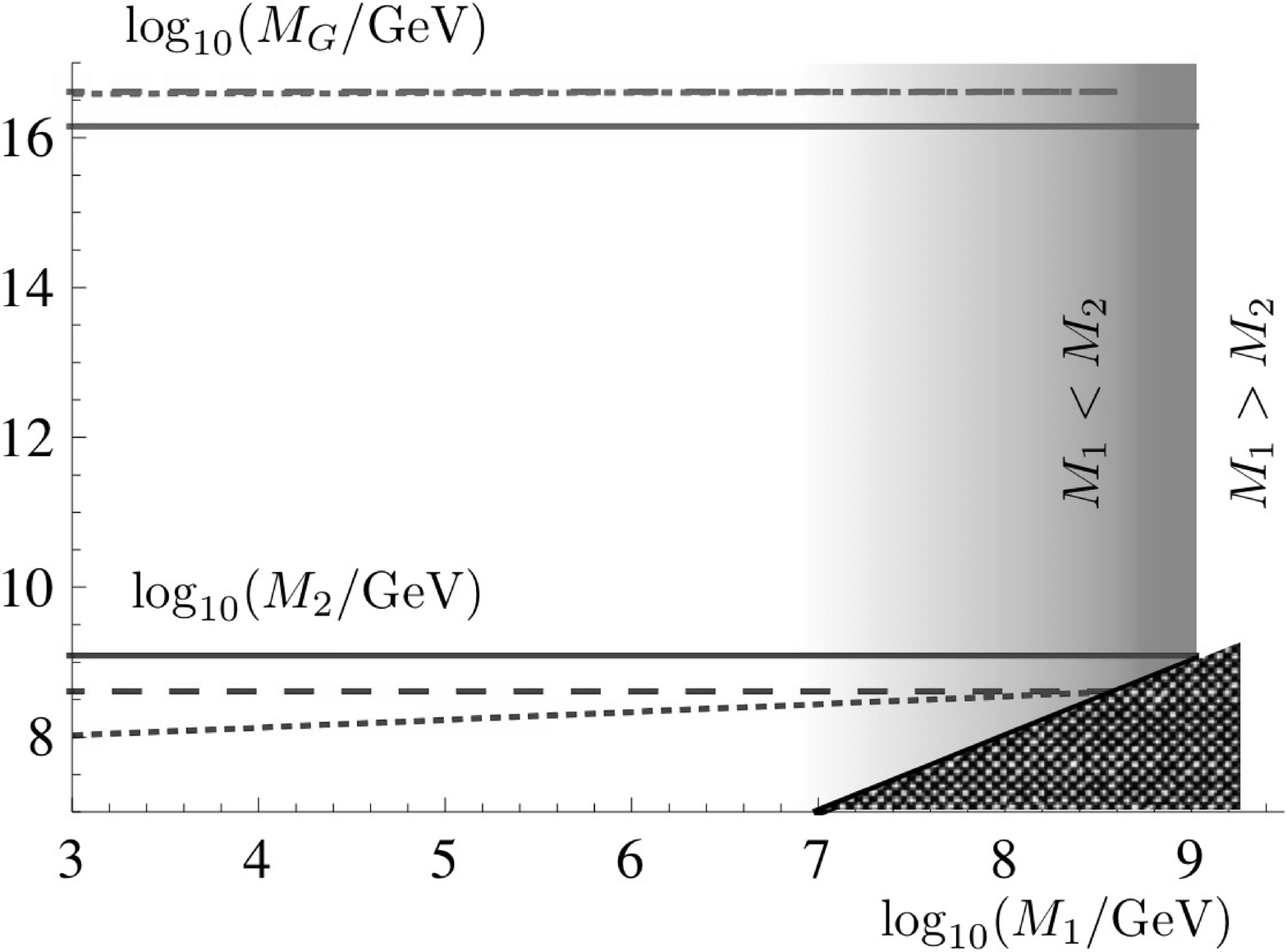}\hskip 20mm
 \includegraphics[width=6cm, height=5cm]{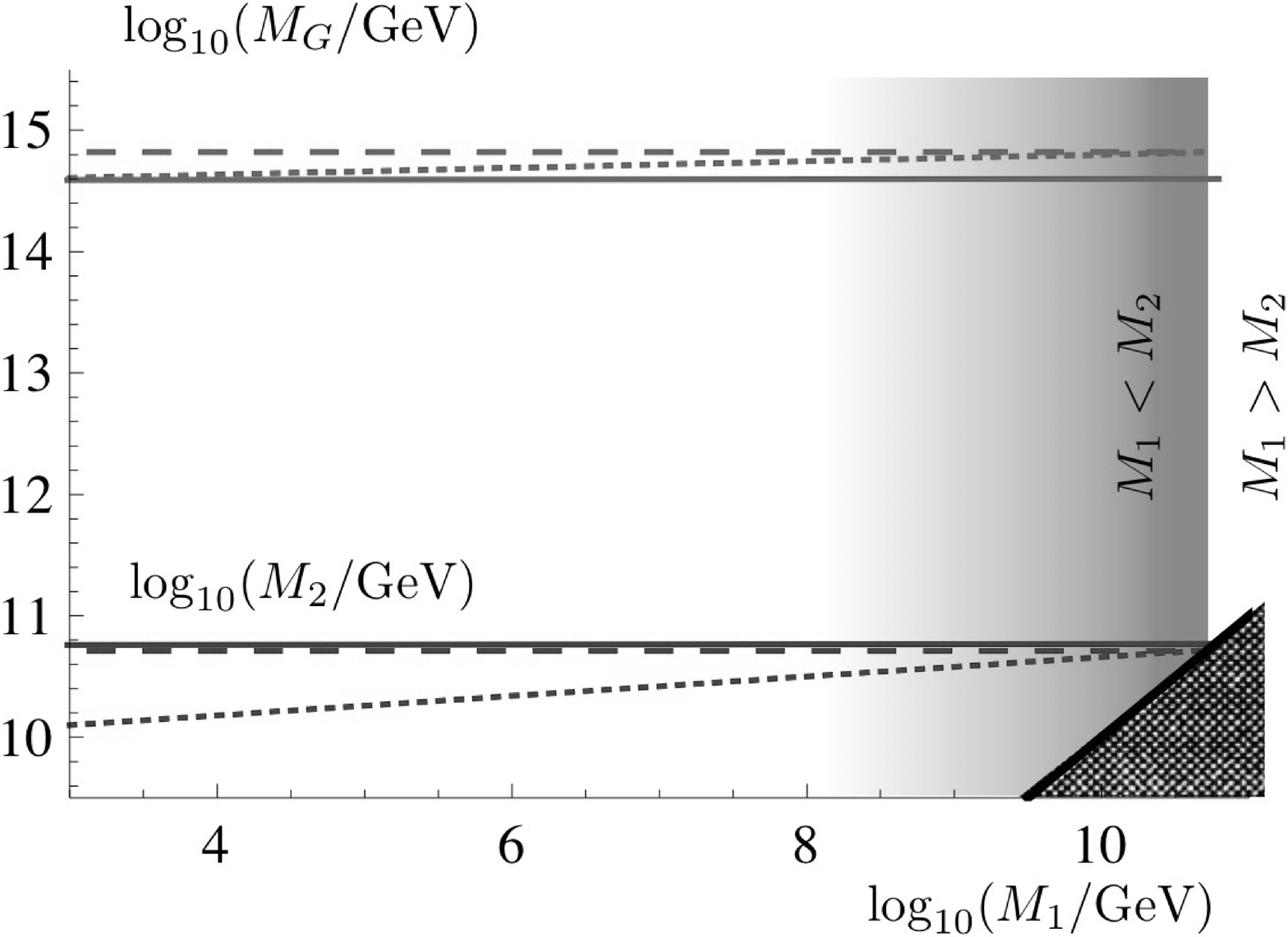}
 \caption{\label{fig:minimalsurvival}Constraints on the GUT scale $M_{G}$ and intermediate scales ($M_{2}$, $M_{1}$)  in the minimal renormalizable $SO(10)$ GUT derived under the assumption of minimal survival~\cite{delAguila:1980at}. The left panel depicts the situation in the $SO(10)\stackrel{M_{G}}{\rightarrow}
SU(3)_c \otimes SU(2)_L \otimes SU(2)_R\otimes U(1)_{B-L} \stackrel{M_{2}}{\rightarrow}SU(3)_c \otimes SU(2)_L \otimes U(1)_R\otimes U(1)_{B-L} \stackrel{M_{1}}{\rightarrow} 
\mbox{SM}$ symmetry breaking chains, the right panel gives the same for the 
$SO(10)\stackrel{M_{G}}{\rightarrow}
SU(4)_C \otimes SU(2)_L \otimes U(1)_R \stackrel{M_{2}}{\rightarrow}SU(3)_c \otimes SU(2)_L \otimes U(1)_R\otimes U(1)_{B-L} \stackrel{M_{1}}{\rightarrow} 
\mbox{SM}$ descents. The shaded area defines the frontier between consistent $(M_{1}<M_{2})$ and inconsistent $(M_{1}>M_{2})$ settings. The three different types of horizontal curves (dotted, dashed, solid) correspond, consecutively, to a one-loop analysis without $U(1)_{R}\otimes U(1)_{B-L}$ mixing effects taken into account,  full-featured one-loop approach and a full two-loop calculation. The horizontality of the latter two can be justified by a simple diagrammatic argument, see, for instance, \cite{Fonseca:2011vn}.}
\end{figure}

Without any detailed information about the scalar spectrum of a theory under consideration, the best one can do in order to study the relevant gauge coupling unification patterns is to employ the minimum survival hypothesis (MSH)~\cite{delAguila:1980at}, i.e., to assume that the components of the unified-theory multiplets cluster around the specific symmetry-breaking scales. As rough as this approximation sounds, it often gives a qualitatively good first look at the salient features of the unification pattern. 
In the non-SUSY $SO(10)$ framework, the ``natural'' positions of the seesaw and grand unification scales have been, under this assumption, studied in \cite{Chang:1984qr,Deshpande:1992em,Deshpande:1992au} and later  reviewed in \cite{Bertolini:2009qj}. 
\vskip 1mm
In order to retain a grip on neutrinos and keep the theory well under control, in what follows we shall focus entirely on the $45\oplus 126$  realization of the  Higgs  mechanism in the minimal $SO(10)$ GUT scheme in which the type-I+II seesaw mechanism is supported at the renormalizable level. 
For more information about this framework an interested reader is referred to the relevant literature~\cite{Bertolini:2012im}.
The resulting constraints on the unification and intermediate scales obtained in this scenario (in the minimal-survival approximation) are displayed in FIGURE~\ref{fig:minimalsurvival}.

Remarkably enough, for both descends of interest there turn out to be  stringent upper limits on the seesaw scale in the minimal survival picture well below $10^{11}$ GeV and, moreover, for the chains passing through the intermediate $SU(4)_C \otimes SU(2)_L \otimes U(1)_R$ stage, the upper limit for $M_{G}$ is in the region which tends to be problematic from the proton lifetime perspective.  This, however, implies that seesaw scale is far outside the  $10^{12-14}$ GeV domain favoured by the light neutrino masses unless the Dirac neutrino mass terms are artificially suppressed. Although there is nothing a-priori wrong about this option we shall not entertain it here. 

Rather than that, we shall attempt to do better than the naive estimates above by exploiting the main drawbacks of the minimum survival approach: First, the MSH does not reflect many important features of realistic spectral patterns (such as, e.g., splitting among different components of multiplets below the relevant symmetry-breaking scales). Second, it is totally ignorant of special regions of the parameter space where the scalar spectrum exhibits {\it unexpected} features such as, e.g., accidentally light states deep in the desert. However, these are exactly the cases when the unification picture can be altered considerably.

\subsection{Consistency beyond minimal survival}
Beyond the minimum-survival approximation, the only guiding principle left for an adventurous parameter-space explorer is the overall consistency of the theory. 
This has several basic aspects:
\paragraph{Non-tachyonic scalar spectrum} First, all potentially interesting regions of the parameter space should support stable (or at least metastable) vacua.  Since the full-featured vacuum stability analysis is very difficult, we shall stick only to the necessary condition, i.e.,  that there should be no tachyonic states in the scalar spectrum. Let us point out that, for each such vacuum configuration at hand, one can obtain other viable settings by, e.g., rescaling all dimensionful parameters in the scalar potential by a common factor. Similarly, it is clear that fiddling around with the mass of an accidentally light state within a range well below the typical mass-scale of all other heavy states does not destabilize specific vacua either because such variations correspond to only very small shifts in the fundamental parameters of the theory.  These two ``degrees of freedom'' can subsequently be used as an efficient tool for reducing the complexity of the numerical analysis of consistent unification patterns. 

\paragraph{Current proton decay limits}
Another obvious constraint on the parameter space of the minimal $SO(10)$ GUT comes from the proton decay; in particular, the current best limit for the $p\to e^{+}\pi^{0}$ mode from Super-Kamiokande~\cite{Nakamura:2010zzi} should be accommodated. In what follows we shall use this together with two assumed future limits that Hyper-Kamiokande (HK)~\cite{Abe:2011ts} may reach by 2025 and 2040 (if built):
\begin{equation}\label{protondecay}
\tau({\rm SK, 2011})> 8.2 \times 10^{33}\, {\rm years}\,,\quad\tau({\rm HK, 2025}) >   9 \times 10^{34}\, {\rm years}\,,\quad
\tau({\rm HK, 2040}) >  2 \times 10^{35}\, {\rm years}\,.
\end{equation}
Furthermore, we shall for simplicity neglect all the details related to the flavour structure of the baryon-number-violating currents so that the numbers above translate directly to the bounds on the position of the GUT scale. In the relevant plots (namely, FIGUREs~\ref{FigureRemnants} and \ref{Figure2}), the points falling between these limits will be distinguished by a simple colour-code where the light grey is used for proton lifetimes between  $8.2 \times 10^{33}$ and $9 \times 10^{34}$ years, dark grey corresponds to lifetimes between $9 \times 10^{34}$ and $2 \times 10^{35}$ years and  black points yield more than $2 \times 10^{35}$ years. 

\paragraph{Big-bang nucleosynthesis} Third, accidentally light coloured states should not be too-long-lived otherwise their late decays may interfere with the highly successful classical  big-bang-nucleosynthesis (BBN) account of the light elements' abundances. Actually, as we shall see, this is not a problem here because the accidentally light multiplets in all fully consistent cases originate in $126_{H}$ and, thus, couple directly to the SM matter fields through the same Yukawa couplings that give rise to, e.g, right-handed neutrino masses. Thus, all the light remnants should decay well before the BBN epoch.

\begin{figure}[t]
 \includegraphics[width=8cm, height=5.3cm]{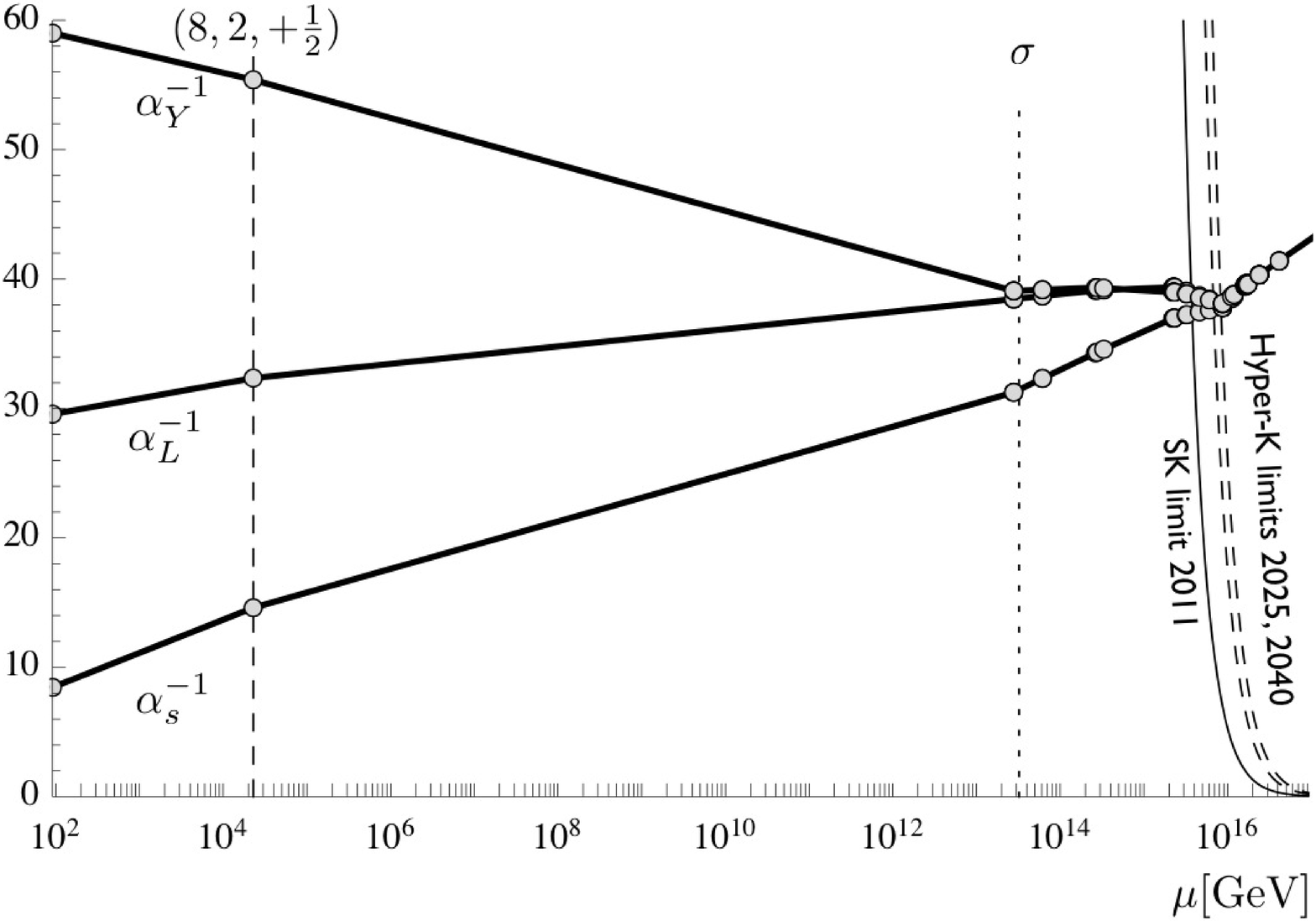}\hskip 5mm
 \includegraphics[width=8cm, height=5.3cm]{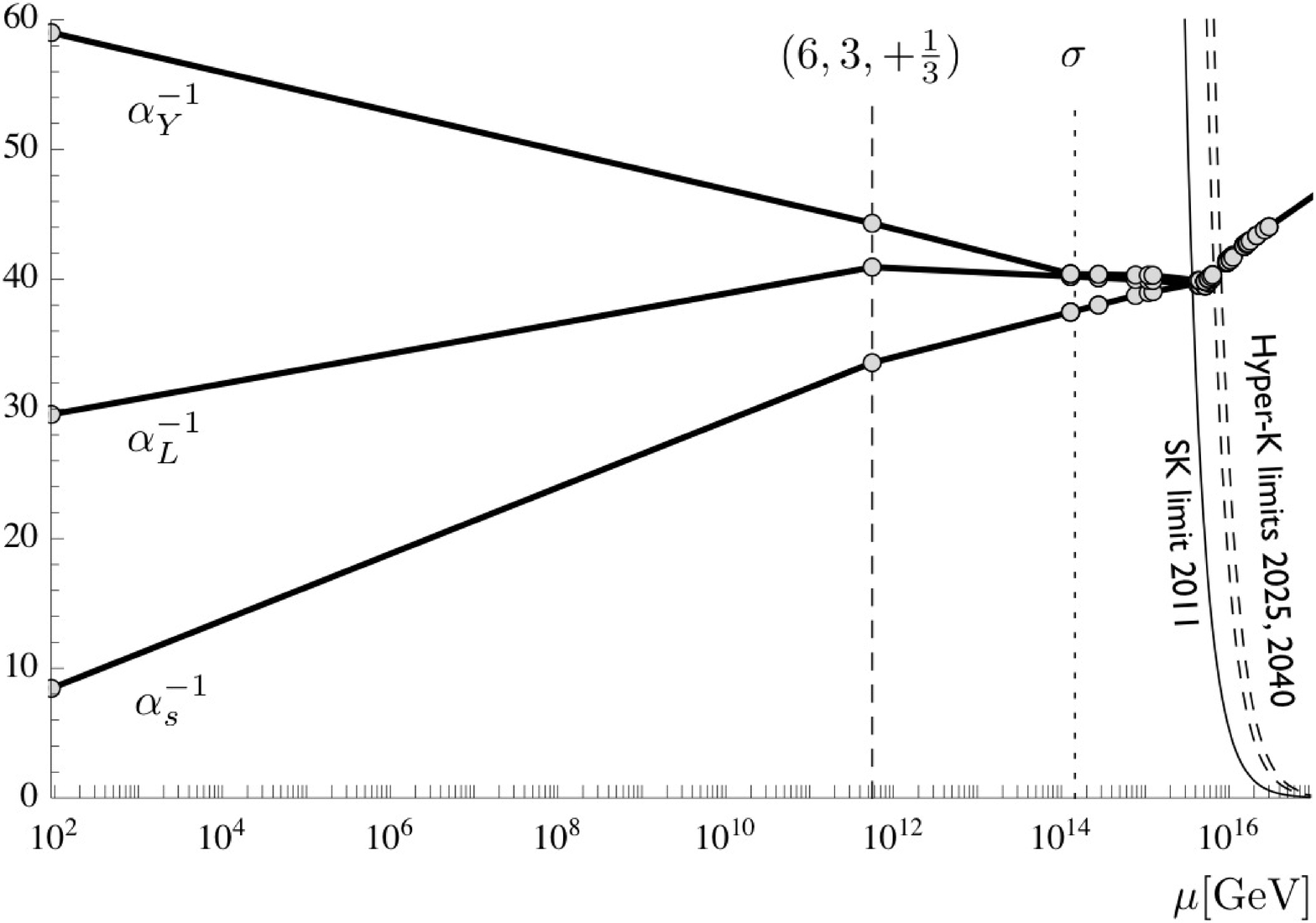}
 \caption{\label{running}The one-loop gauge unification patterns in the two cases of our main interest with the light $(8,2,+\tfrac{1}{2})$ [left panel] or the light $(6,3,+\tfrac{1}{3})$  [right panel]  multiplet in the GUT desert. The almost vertical curves on the right correspond to the naive bounds on the  ``unification point'' position derived from the current Super-K and assumed future Hyper-K limits quoted in eq.~(\ref{protondecay}). There are only three trajectories drawn up to the GUT scale because the calculation has been conveniently performed in the effective SM picture. The gray circles depict positions of various heavy scalar and gauge multiplets (as listed in TABLE~\ref{TableSpectra}) supporting these patterns. For more details an interested reader is referred to the original work~\cite{Bertolini:2012im}.}
\end{figure}

\begin{figure}[t]
 \includegraphics[width=8cm, height=5.3cm]{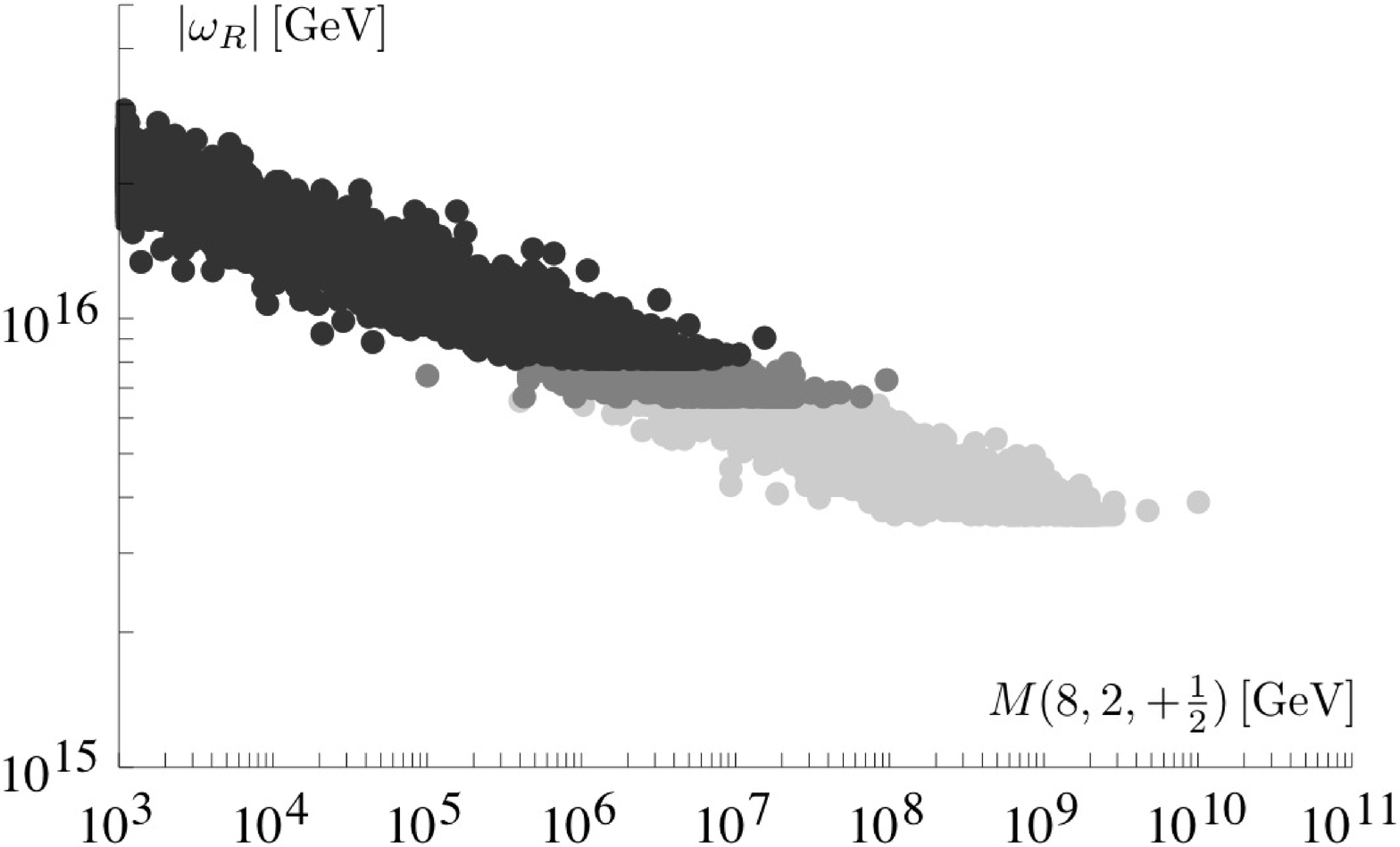}
 \includegraphics[width=8cm, height=5.3cm]{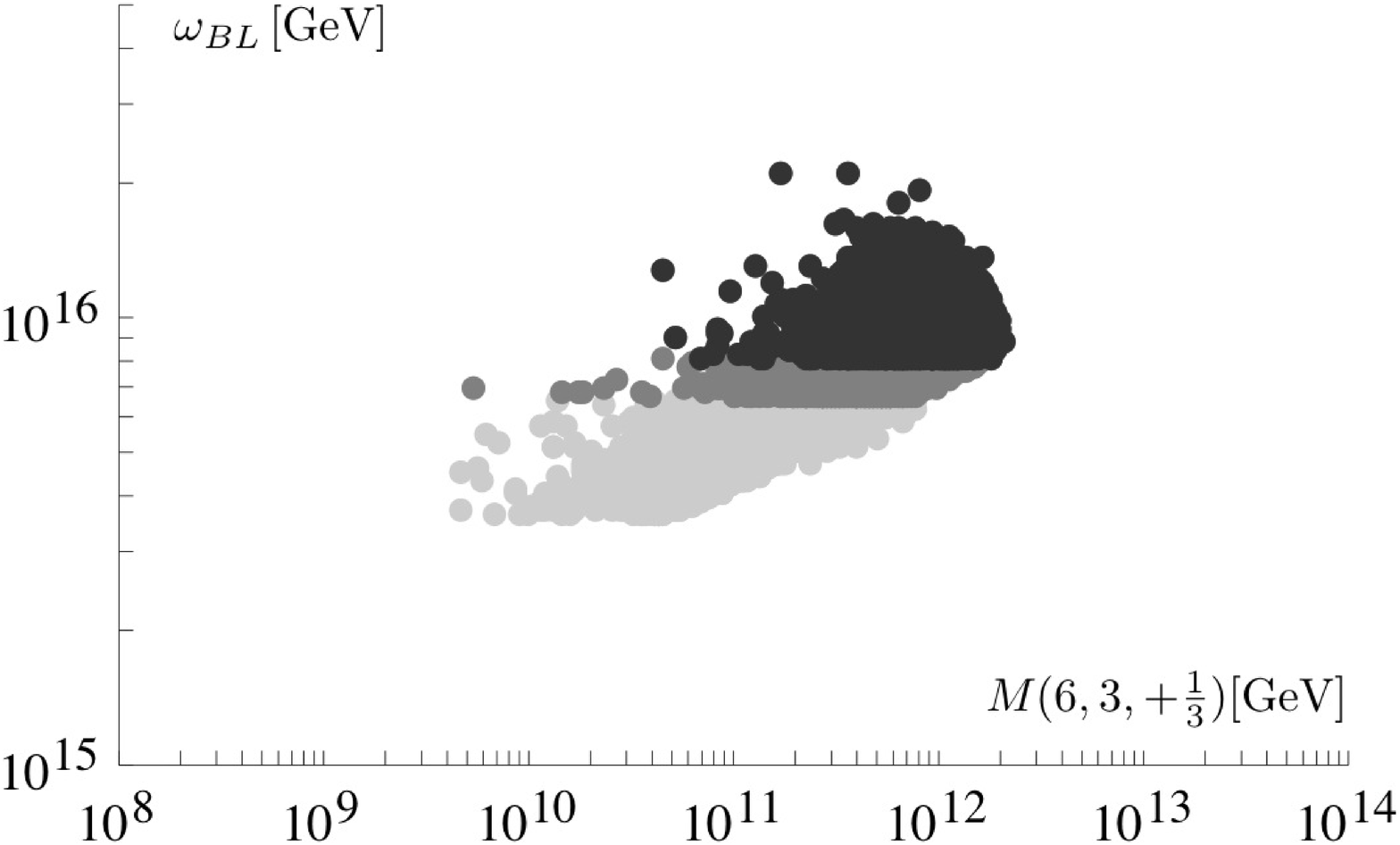}
\caption{\label{FigureRemnants}Mass-ranges for the accidentally light states in the light-octet [left panel] and light sextet [right panel] scenarios. $M(6,3,+\tfrac{1}{3})-\omega_{BL}$. In both cases, the variable on the vertical axis plays a role of the GUT scale; this is why the proton lifetime limits cut the points from below. Remarkably enough, the octet can be rather light and, in principle, it can be pulled close to the electroweak scale where it can, e.g., leave its imprints in the LHC searches. }
\end{figure}

\paragraph{Consistent unification patterns}
The simple constraints above are enough to filter out all but two qualitatively different settings with a single accidentally light scalar multiplet well below the $B-L$ symmetry-breaking scale: a scenario with a very light colour octet $(8,2,+\tfrac{1}{2})$ and another scheme with an intermediate-mass-scale colour sextet $(6,3,+\tfrac{1}{3})$.    
The typical shapes of the gauge unification patterns in these two cases are shown in FIGURE~\ref{running}. The results of a detailed numerical scan over extended regions supporting these solutions are given in FIGURE~\ref{FigureRemnants}. Interestingly, the mass range of the octet solution (on the left panel in FIGURE~\ref{FigureRemnants}) can stretch as low as to the TeV domain so, in principle, it can even leave its imprints in the LHC searches; however, the sextet is not allowed  below roughly $10^{9}$~GeV.

\subsection{Seesaw scale upper limits in consistent scenarios}
\begin{figure}[t]
 \includegraphics[width=8cm, height=5.3cm]{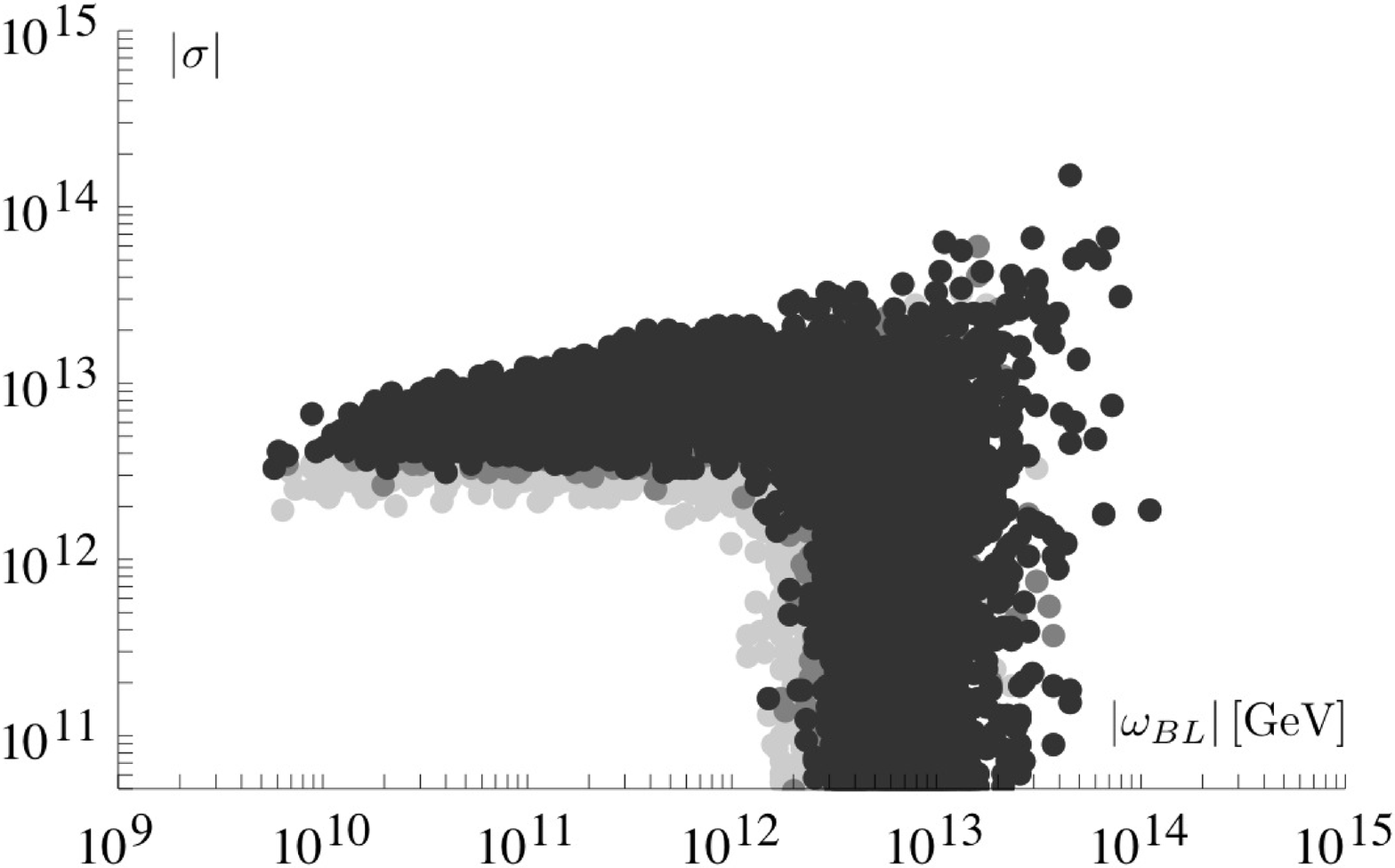}
 \includegraphics[width=8cm, height=5.3cm]{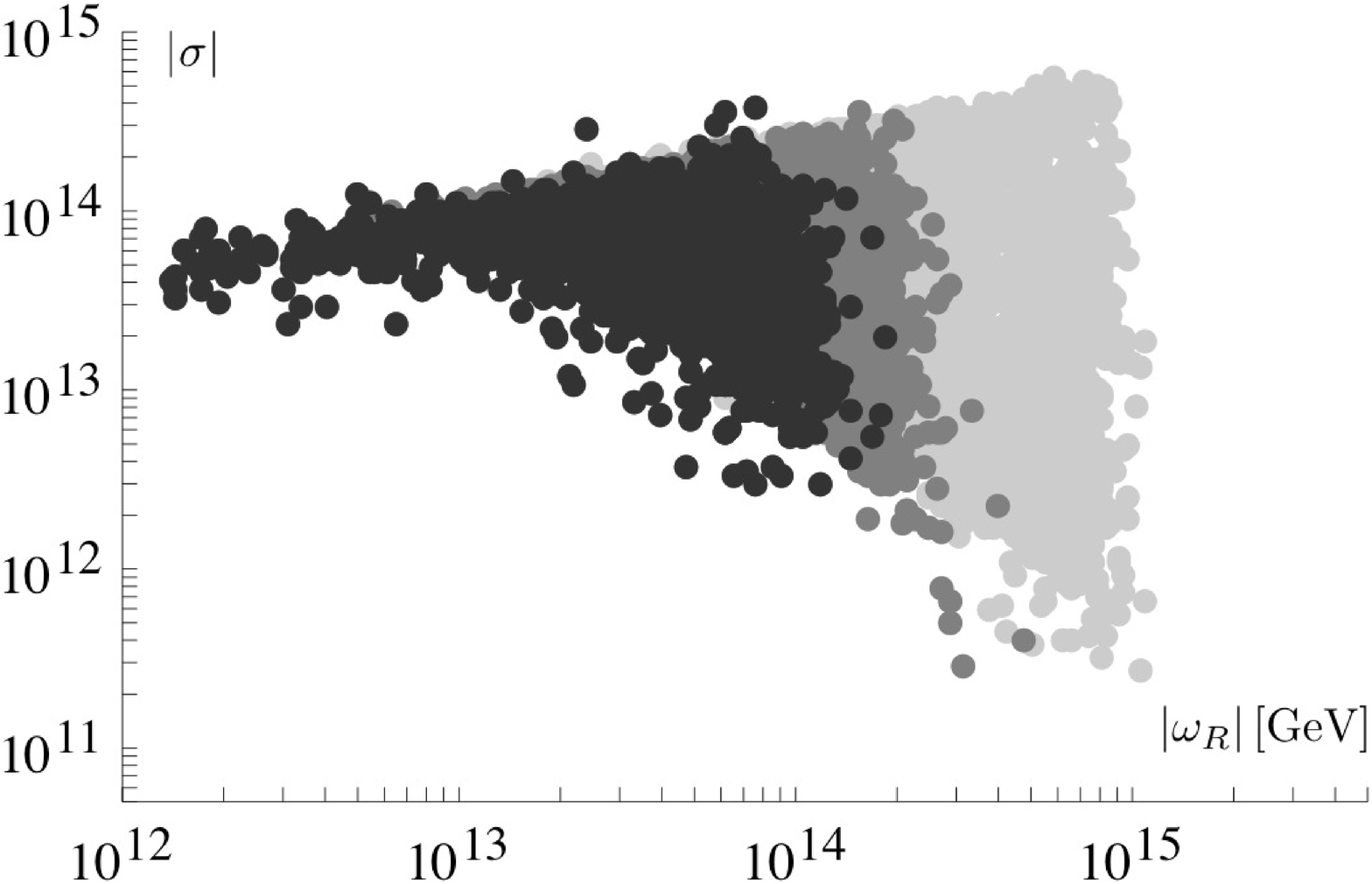}
 \caption{\label{Figure2}The $|\omega_{R}|-|\sigma|$ and  $|\omega_{BL}|-|\sigma|$ cuts of the parameter space corresponding to solutions with the light $(8,2,+\tfrac{1}{2})$ [left panel] or $(6,3,+\tfrac{1}{3})$  [right panel]  multiplets in the desert where $\sigma$ denotes the $B-L$-breaking VEV of $126_{H}$. Various levels of gray correspond to domains accessible for different GUT-scale limits, cf.~(\ref{protondecay}).}
\end{figure}

Finally, the allowed ranges for the $B-L$ -breaking VEV (denoted $\sigma$) in these two scenarios are depicted in FIGURE~\ref{Figure2}. Remarkably enough, {\it the naive MSH-based upper bounds on the seesaw scale are in both cases relaxed by as much as four orders of magnitude} as they stretch up to about $10^{14}$ GeV in the light-octet case and up to almost $5\times 10^{14}$ GeV in the case of the light sextet. This, however, makes the implementation of the standard seesaw mechanism possible even without resorting to the excessive fine-tuning in the Yukawa sector implied in  previous studies.

\section{Conclusions and outlook}
Even though the minimal SO(10) models have been recently revived as consistent quantum field theories free of inherent tachyonic instabilities, this beautiful and simple framework has never been rehabilitated as a potentially realistic and predictive GUT scheme. This was namely due to the old studies of the relevant gauge unification patterns which revealed a generic tendency for the $B-L$ symmetry breaking scale to be confined below $10^{11}$ GeV, apparently too low for a reasonable implementation of the seesaw mechanism for neutrino masses. However, all these early studies based on the minimal survival hypothesis suffer from a generic incapability to account for the effects of accidentally light multiplets with masses detached from any specific symmetry breaking scale. As we argued, a closer inspection of the unification constraints reveals a much wider room for the $B-L$ breaking VEV stretching up to the $10^{14}$ GeV ballpark which, in turn, allows the seesaw picture to be implemented without excessive fine-tuning.  

Besides that, the minimal renormalizable SO(10) scenario has several other interesting features which make it an interesting candidate for a further theoretical scrutiny:
\paragraph{Possible LHC imprints of the light octet scenario}
 Remarkably enough, the same pair of scalars that we identified as powerful ``running helpers'' in the minimal SO(10) framework was recently singled out in the work \cite{Dorsner:2012pp} from a totally different perspective. There,  the apparent enhancement in the $H\to \gamma \gamma$ rate indicated by the current LHC data was shown to be attributable to just this couple of states if any of them falls into the vicinity of the TeV scale.
However, at the current level of accuracy only the octet $(8,2,+\tfrac{1}{2})$ can be light enough in the minimal $SO(10)$ to play any role in the Higgs physics because the sextet is not allowed below about $10^{9}$~GeV, see FIGURE~\ref{FigureRemnants}. Nevertheless, the proton-decay limits used in cutting the low-mass-sextet region therein are rather naive and it can happen that a more-sophisticated analysis including flavour effects opens a bigger room for the light sextet too.

\paragraph{Suppression of the Planck-scale induced unification-smearing effects in the minimal $SO(10)$ GUTs}
Besides simplicity, the $SO(10)$ models in which the GUT-scale symmetry breaking is driven by a VEV of the 45-dimensional adjoint representation have another very interesting feature. This has to do with the general fragility of grand unification with respect to the Planck-scale ($M_{Pl}$) effects which, given the proximity of $M_{G}$ and $M_{Pl}$, may not be entirely negligible. Concerning their possible impact on, e.g., proton lifetime estimates, the most important of these is namely the Planck-scale induced violation of the canonical normalization of the heavy gauge fields~\cite{Calmet:2008df, Chakrabortty:2008zk} due to the higher-order corrections to the gauge kinetic form emerging already at the $d=5$  level: ${\cal L}^{(5)}\ni {\rm Tr} [ F_{\mu\nu}HF^{\mu\nu}]/M_{Pl}$;
here $H$ is any scalar in the theory which can couple to a pair of adjoint representations of a specific GUT symmetry group, i.e., any field appearing in the symmetric part of the decomposition of their tensor product. For a GUT-scale VEV of $H$, this induces a percent-level effect which, after a suitable redefinition of the gauge fields, leads to similar-size shifts in the GUT-scale matching conditions. Such a ``unification smearing effect'' can, subsequently, play a significant role in an accurate NLO GUT-scale determination which, in turn, further adds to the existing theoretical uncertainties in the absolute proton lifetime estimates. 

However, if in the SO(10) the GUT-scale symmetry breaking is triggered by the VEV of $45_{H}$, this problem is absent because 
${\rm Tr} [ F_{\mu\nu}45_{H}F^{\mu\nu}]=0$ due to the fact that 45 is not in the symmetric part of the $45\otimes 45$ decomposition [recall that $(45\otimes 45)_{\rm sym}=54\oplus 210\oplus 770$]. Thus, the minimal $SO(10)$ scheme with the adjoint-driven Higgs mechanism is uniquely robust with respect to this class of quantum gravity effects. This makes the symmetry-breaking analysis more reliable and, hence, admits in principle for a strong reduction of this type of  theoretical uncertainties in the proton lifetime estimates.   

\paragraph{Proton lifetime at the next-to-leading-order level}
The simplicity of the minimal $SO(10)$ scenario advocated in this study, together with its rather unique robustness with respect to the Planck-scale-induced unification smearing effects make this class of models particularly suitable for a possible next-to-leading-order (NLO) proton lifetime analysis. That, however, is far from trivial. 
To this end, let us just note that the main source of the large theoretical uncertainties in the existing proton lifetime estimates, cf. FIGURE~\ref{estimates}, is the inaccuracy of the GUT-scale determination, partly  due to the uncertainties in the low-energy  inputs (especially in $\alpha_{s}(M_{Z})$) and, in particular, the limited precision of the one-loop approach -- given the logarithmic nature of the renormalization-group evolution, both these errors are exponentially amplified in the resulting proton decay amplitude. The only way to keep such uncertainties  under control  is thus a careful two-loop renormalization-group calculation including, as a prerequisite, the one-loop spectrum of the theory resulting from a dedicated analysis (like, e.g., that in \cite{Bertolini:2009es}) together with the proper one-loop matching \cite{Weinberg:1980wa,Hall:1980kf} conditions. In this respect, the minimal renormalizable $SO(10)$ GUT of our main concern here can be {\it the} scenario in which a decisive NLO proton lifetime analysis can be just at the verge of tractability.   

\paragraph{GUT-scale baryogenesis}
 The option of a GUT-scale baryogenesis in SO(10) models, recently revived in \cite{Babu:2012vb}, is another interesting feature of the model under consideration. On the technical side, however, the amount of thus generated baryon asymmetry depends on the size of the quartic coupling $\eta_{2}$ in the scalar potential, cf. formula (3) in reference~\cite{Bertolini:2012im}, which, unfortunately, turns out to be one of the most elusive of all the theory parameters -- as it was argued in \cite{Bertolini:2012im}, $\eta_{2}$ does not enter the tree-level scalar spectrum and, so, the one-loop unification pattern does not impose any constraints on it.   
However, this coupling does take part in the decomposition of the light Higgs boson in terms of its defining components in $10_{H}\oplus 126_{H}$ and, thus, it may be constrained indirectly by the flavour structure of the effective theory.  Hence, a dedicated proton lifetime analysis advocated above (which, inevitably, must include a detailed account of the flavour structure of the model) may, as one of its by-products, provide also a better grip on the GUT-scale baryogenesis in the minimal $SO(10)$ GUT.

\subsection{Acknowledgments}
M.M. is grateful for the invitation to CETUP'12 and support during his stay in Lead, SD.   
S.B. is partially supported by MIUR and the EU UNILHC-grant agreement PITN-GA-2009-237920. The work of L.DL. was supported by the DFG through the SFB/TR 
9 ``Computational Particle Physics''. The work of M.M. is supported by the Marie Curie Intra European Fellowship
within the 7th European Community Framework Programme
FP7-PEOPLE-2009-IEF, contract number PIEF-GA-2009-253119, by the EU
Network grant UNILHC PITN-GA-2009-237920, by the Spanish MICINN
grants FPA2008-00319/FPA and MULTIDARK CAD2009-00064
(Consolider-Ingenio 2010 Programme) and by the Generalitat
Valenciana grant Prometeo/2009/091.

\renewcommand{\arraystretch}{1.4}
\begin{table}
\begin{tabular}{c|c|c|c}
\hline
multiplet & type  & $\Delta b^{321}$ & mass [GeV]  \\
\hline
$\bf (6,3,+\tfrac{1}{3})$ & {\bf CS} & $(\tfrac{5}{2}, 4, \tfrac{2}{5})$ & $\bf 5.6 \times 10^{11}$ \\
$(1,1,-1)$ & VB & $(0, 0, -\tfrac{11}{5})$ & $1.3 \times 10^{14}$ \\
$(1,1,+1)$ & VB & $(0, 0, -\tfrac{11}{5})$ & $1.3 \times 10^{14}$ \\
$(1,1,+1)$ & GB & $(0, 0, \tfrac{1}{5})$ & $1.3 \times 10^{14}$ \\
$(1,1,0)$ & VB & $(0, 0, 0)$ & $2.8 \times 10^{14}$ \\
$(1,1,0)$ & GB & $(0, 0, 0)$ & $2.8 \times 10^{14}$ \\
$(8,1,0)$ & RS & $(\tfrac{1}{2}, 0, 0)$ & $7.7 \times 10^{14}$ \\
$(3,2,+\tfrac{1}{6})$ & CS & $(\tfrac{1}{3}, \tfrac{1}{2}, \tfrac{1}{30})$ & $1.1 \times 10^{15}$ \\
$(3,2,+\tfrac{7}{6})$ & CS & $(\tfrac{1}{3}, \tfrac{1}{2}, \tfrac{49}{30})$ & $1.2 \times 10^{15}$ \\
$(1,1,0)$ & RS & $(0, 0, 0)$ & $4.3 \times 10^{15}$ \\
$(1,1,+2)$ & CS & $(0, 0, \tfrac{4}{5})$ & $4.5 \times 10^{15}$ \\
$\bf(\overline{3},2,-\tfrac{1}{6})$ & \bf VB & $(-\tfrac{11}{3}, -\tfrac{11}{2}, -\tfrac{11}{30})$ & $\bf5.2 \times 10^{15}$ \\
$\bf(3,2,+\tfrac{1}{6})$ & \bf VB & $(-\tfrac{11}{3}, -\tfrac{11}{2}, -\tfrac{11}{30})$ & $\bf5.2 \times 10^{15}$ \\
$\bf (3,2,+\tfrac{1}{6})$ & \bf GB & $(\tfrac{1}{3}, \tfrac{1}{2}, \tfrac{1}{30})$ & $\bf 5.2 \times 10^{15}$ \\
$\bf(\overline{3},2,+\tfrac{5}{6})$ & \bf VB & $(-\tfrac{11}{3}, -\tfrac{11}{2}, -\tfrac{55}{6})$ & $\bf5.2 \times 10^{15}$ \\
$\bf(3,2,-\tfrac{5}{6})$ & \bf VB & $(-\tfrac{11}{3}, -\tfrac{11}{2}, -\tfrac{55}{6})$ & $\bf5.2 \times 10^{15}$ \\
$\bf (3,2,-\tfrac{5}{6})$ & \bf GB & $(\tfrac{1}{3}, \tfrac{1}{2}, \tfrac{5}{6})$ & $\bf 5.2 \times 10^{15}$ \\
$(1,1,+1)$ & CS & $(0, 0, \tfrac{1}{5})$ & $5.6 \times 10^{15}$ \\
$(1,1,0)$ & RS & $(0, 0, 0)$ & $5.7 \times 10^{15}$ \\
$(1,3,0)$ & RS & $(0, \tfrac{1}{3}, 0)$ & $6.1 \times 10^{15}$ \\
$(\overline{3},1,+\tfrac{1}{3})$ & CS & $(\tfrac{1}{6}, 0, \tfrac{1}{15})$ & $6.4 \times 10^{15}$ \\
$(8,2,+\tfrac{1}{2})$ & CS & $(2, \tfrac{4}{3}, \tfrac{4}{5})$ & $9.3 \times 10^{15}$ \\
$(\overline{3},1,+\tfrac{4}{3})$ & CS & $(\tfrac{1}{6}, 0, \tfrac{16}{15})$ & $9.6 \times 10^{15}$ \\
$(\overline{3},1,+\tfrac{1}{3})$ & CS & $(\tfrac{1}{6}, 0, \tfrac{1}{15})$ & $9.6 \times 10^{15}$ \\
$(\overline{3},1,-\tfrac{2}{3})$ & CS & $(\tfrac{1}{6}, 0, \tfrac{4}{15})$ & $9.6 \times 10^{15}$ \\
$(\overline{3},1,-\tfrac{2}{3})$ & VB & $(-\tfrac{11}{6}, 0, -\tfrac{44}{15})$ & $1.0 \times 10^{16}$ \\
$(3,1,+\tfrac{2}{3})$ & VB & $(-\tfrac{11}{6}, 0, -\tfrac{44}{15})$ & $1.0 \times 10^{16}$ \\
$(\overline{3},1,-\tfrac{2}{3})$ & GB & $(\tfrac{1}{6}, 0, \tfrac{4}{15})$ & $1.0 \times 10^{16}$ \\
$(8,2,+\tfrac{1}{2})$ & CS & $(2, \tfrac{4}{3}, \tfrac{4}{5})$ & $1.1 \times 10^{16}$ \\
$(\overline{6},1,+\tfrac{2}{3})$ & CS & $(\tfrac{5}{6}, 0, \tfrac{8}{15})$ & $1.5 \times 10^{16}$ \\
$(1,2,+\tfrac{1}{2})$ & {RS} & $(0, \tfrac{1}{12}, \tfrac{1}{20})$ & $1.5 \times 10^{16}$ \\
$(\overline{6},1,-\tfrac{1}{3})$ & CS & $(\tfrac{5}{6}, 0, \tfrac{2}{15})$ & $1.5 \times 10^{16}$ \\
$(\overline{6},1,-\tfrac{4}{3})$ & CS & $(\tfrac{5}{6}, 0, \tfrac{32}{15})$ & $1.5 \times 10^{16}$ \\
$(1,2,+\tfrac{1}{2})$ & {RS} & $(0, \tfrac{1}{12}, \tfrac{1}{20})$ & $1.6 \times 10^{16}$ \\
$(\overline{3},1,+\tfrac{1}{3})$ & CS & $(\tfrac{1}{6}, 0, \tfrac{1}{15})$ & $1.7 \times 10^{16}$ \\
$(3,3,-\tfrac{1}{3})$ & CS & $(\tfrac{1}{2}, 2, \tfrac{1}{5})$ & $1.8 \times 10^{16}$ \\
$(3,2,+\tfrac{1}{6})$ & CS & $(\tfrac{1}{3}, \tfrac{1}{2}, \tfrac{1}{30})$ & $2.1 \times 10^{16}$ \\
$(3,2,+\tfrac{7}{6})$ & CS & $(\tfrac{1}{3}, \tfrac{1}{2}, \tfrac{49}{30})$ & $2.1 \times 10^{16}$ \\
$(1,3,-1)$ & CS & $(0, \tfrac{2}{3}, \tfrac{3}{5})$ & $2.6 \times 10^{16}$ \\
$(1,1,0)$ & RS & $(0, 0, 0)$ & $3.0 \times 10^{16}$ \\
\hline
\end{tabular}
\;\;\;\;\;\;
\begin{tabular}{c|c|c|c}
\hline
multiplet & type   & $\Delta b^{321}$ & mass [GeV]  \\
\hline
$\bf (8,2,+\tfrac{1}{2})$ & {\bf CS} & $(2, \tfrac{4}{3}, \tfrac{4}{5})$ & $\bf 2.3 \times 10^{4}$ \\
$(\overline{3},1,-\tfrac{2}{3})$ & VB & $(-\tfrac{11}{6}, 0, -\tfrac{44}{15})$ & $2.8 \times 10^{13}$ \\
$(3,1,+\tfrac{2}{3})$ & VB & $(-\tfrac{11}{6}, 0, -\tfrac{44}{15})$ & $2.8 \times 10^{13}$ \\
$(\overline{3},1,-\tfrac{2}{3})$ & GB & $(\tfrac{1}{6}, 0, \tfrac{4}{15})$ & $2.8 \times 10^{13}$ \\
$(1,1,0)$ & VB & $(0, 0, 0)$ & $6.1 \times 10^{13}$ \\
$(1,1,0)$ & GB & $(0, 0, 0)$ & $6.1 \times 10^{13}$ \\
$(3,2,+\tfrac{7}{6})$ & CS & $(\tfrac{1}{3}, \tfrac{1}{2}, \tfrac{49}{30})$ & $2.6 \times 10^{14}$ \\
$(3,2,+\tfrac{1}{6})$ & CS & $(\tfrac{1}{3}, \tfrac{1}{2}, \tfrac{1}{30})$ & $2.8 \times 10^{14}$ \\
$(1,2,+\tfrac{1}{2})$ & {RS} & $(0, \tfrac{1}{12}, \tfrac{1}{20})$ & $3.3 \times 10^{14}$ \\
$(1,1,0)$ & RS & $(0, 0, 0)$ & $2.2 \times 10^{15}$ \\
$(\overline{3},1,-\tfrac{2}{3})$ & CS & $(\tfrac{1}{6}, 0, \tfrac{4}{15})$ & $2.3 \times 10^{15}$ \\
$(6,3,+\tfrac{1}{3})$ & CS & $(\tfrac{5}{2}, 4, \tfrac{2}{5})$ & $2.3 \times 10^{15}$ \\
$(3,3,-\tfrac{1}{3})$ & CS & $(\tfrac{1}{2}, 2, \tfrac{1}{5})$ & $2.3 \times 10^{15}$ \\
$(1,3,-1)$ & CS & $(0, \tfrac{2}{3}, \tfrac{3}{5})$ & $2.3 \times 10^{15}$ \\
$(\overline{6},1,-\tfrac{4}{3})$ & CS & $(\tfrac{5}{6}, 0, \tfrac{32}{15})$ & $3.2 \times 10^{15}$ \\
$(1,1,0)$ & RS & $(0, 0, 0)$ & $3.3 \times 10^{15}$ \\
$(8,1,0)$ & RS & $(\tfrac{1}{2}, 0, 0)$ & $4.6 \times 10^{15}$ \\
$(1,3,0)$ & RS & $(0, \tfrac{1}{3}, 0)$ & $6.1 \times 10^{15}$ \\
$\bf(\overline{3},2,+\tfrac{5}{6})$ & \bf VB & $(-\tfrac{11}{3}, -\tfrac{11}{2}, -\tfrac{55}{6})$ & $\bf8.7 \times 10^{15}$ \\
$\bf(3,2,-\tfrac{5}{6})$ & \bf VB & $(-\tfrac{11}{3}, -\tfrac{11}{2}, -\tfrac{55}{6})$ & $\bf8.7 \times 10^{15}$ \\
$\bf (3,2,-\tfrac{5}{6})$ & \bf GB & $(\tfrac{1}{3}, \tfrac{1}{2}, \tfrac{5}{6})$ & $\bf 8.7 \times 10^{15}$ \\
$\bf(\overline{3},2,-\tfrac{1}{6})$ & \bf VB & $(-\tfrac{11}{3}, -\tfrac{11}{2}, -\tfrac{11}{30})$ & $\bf8.7 \times 10^{15}$ \\
$\bf(3,2,+\tfrac{1}{6})$ & \bf VB & $(-\tfrac{11}{3}, -\tfrac{11}{2}, -\tfrac{11}{30})$ & $\bf8.7 \times 10^{15}$ \\
$\bf (3,2,+\tfrac{1}{6})$ & \bf GB & $(\tfrac{1}{3}, \tfrac{1}{2}, \tfrac{1}{30})$ & $\bf 8.7 \times 10^{15}$ \\
$(\overline{3},1,+\tfrac{1}{3})$ & CS & $(\tfrac{1}{6}, 0, \tfrac{1}{15})$ & $1.1 \times 10^{16}$ \\
$(\overline{3},1,+\tfrac{1}{3})$ & CS & $(\tfrac{1}{6}, 0, \tfrac{1}{15})$ & $1.2 \times 10^{16}$ \\
$(1,1,+1)$ & CS & $(0, 0, \tfrac{1}{5})$ & $1.6 \times 10^{16}$ \\
$(\overline{3},1,+\tfrac{1}{3})$ & CS & $(\tfrac{1}{6}, 0, \tfrac{1}{15})$ & $1.6 \times 10^{16}$ \\
$(\overline{6},1,-\tfrac{1}{3})$ & CS & $(\tfrac{5}{6}, 0, \tfrac{2}{15})$ & $1.6 \times 10^{16}$ \\
$(3,2,+\tfrac{7}{6})$ & CS & $(\tfrac{1}{3}, \tfrac{1}{2}, \tfrac{49}{30})$ & $1.7 \times 10^{16}$ \\
$(1,2,+\tfrac{1}{2})$ &{RS} & $(0, \tfrac{1}{12}, \tfrac{1}{20})$ & $1.7 \times 10^{16}$ \\
$(8,2,+\tfrac{1}{2})$ & CS & $(2, \tfrac{4}{3}, \tfrac{4}{5})$ & $1.7 \times 10^{16}$ \\
$(3,2,+\tfrac{1}{6})$ & CS & $(\tfrac{1}{3}, \tfrac{1}{2}, \tfrac{1}{30})$ & $1.7 \times 10^{16}$ \\
$(1,1,-1)$ & VB & $(0, 0, -\tfrac{11}{5})$ & $1.7 \times 10^{16}$ \\
$(1,1,+1)$ & VB & $(0, 0, -\tfrac{11}{5})$ & $1.7 \times 10^{16}$ \\
$(1,1,+1)$ & GB & $(0, 0, \tfrac{1}{5})$ & $1.7 \times 10^{16}$ \\
$(1,1,+2)$ & CS & $(0, 0, \tfrac{4}{5})$ & $2.4 \times 10^{16}$ \\
$(\overline{3},1,+\tfrac{4}{3})$ & CS & $(\tfrac{1}{6}, 0, \tfrac{16}{15})$ & $2.4 \times 10^{16}$ \\
$(\overline{6},1,+\tfrac{2}{3})$ & CS & $(\tfrac{5}{6}, 0, \tfrac{8}{15})$ & $2.4 \times 10^{16}$ \\
$(1,1,0)$ & RS & $(0, 0, 0)$ & $4.1 \times 10^{16}$ \\
\hline
\end{tabular}
\caption{\label{TableSpectra}The scalar spectra underpinning the gauge-coupling evolution in FIGURE \ref{running}. The ``type'' column encodes the nature of the relevant multiplet as follows: CS=complex scalar, RS=real scalar, GB=Goldstone boson and VB=vector boson, $\Delta b^{321}$ corresponds to the changes in the beta-functions across each of the thresholds. The accidentally light scalars (in the first row) and the gauge bosons defining the GUT scale are in boldface. For further details see~\cite{Bertolini:2012im}.}
\label{tab:a}
\end{table}

\end{document}